\documentclass[]{fairmeta_gr2}

\usepackage{amsmath}
\usepackage{amssymb}
\usepackage{tikz}
\usetikzlibrary{shapes,calc}
\usetikzlibrary{arrows.meta,positioning,decorations.pathreplacing,calc}
\definecolor{dgr2blue}{RGB}{52,120,200}

\title{Diffusion-GR2: Diffusion Generative Reasoning Re-ranker}

\author[2,*]{Zhuoxuan Zhang}
\author[2,*]{Kangqi Ni}
\author[2,*]{Yuhang Chen}
\author[1,*,\dagger,\ddagger]{Mingfu Liang}
\author[1]{Xiaohan Wei}
\author[1]{Yunchen Pu}
\author[1]{Fei Tian}
\author[1]{Chonglin Sun}
\author[1]{Frank Shyu}
\author[1]{Adam (Yang) Song}
\author[1]{Sandeep Pandey}
\author[1,\ddagger]{Luke Simon}
\author[2,\ddagger]{Tianlong Chen}
\author[1,\ddagger]{Xi Liu}

\affiliation[1]{Meta AI}
\affiliation[2]{UNC Chapel Hill}

\contribution[*]{Co-first author}
\contribution[\dagger]{Project and Execution Lead}
\contribution[\ddagger]{Joint corresponding author}

\abstract{%
Generative reasoning re-rankers achieve strong recommendation accuracy by emitting a chain-of-thought before re-ordering a candidate list, but they are slow at inference: an autoregressive (AR) decoder spends one sequential forward pass per reasoning token, and the reasoning trace far exceeds the ranking it produces. 
To reduce this cost, block-diffusion language models decode many positions in parallel over a few denoising steps and are substantially faster, yet naively converting an AR re-ranker into one opens two accuracy gaps: (1) a \emph{structural} gap: answer positions are denoised in parallel and scored independently, so the decoder emits invalid rankings (duplicated, dropped, or out-of-set identifiers) that AR avoids through left-to-right masking; 
and (2) a \emph{distributional} gap: fine-tuning the converted model on fixed teacher trajectories is off-policy relative to its own decoding at inference, leaving a residual accuracy gap. To close both gaps while keeping the speedup, we propose \textbf{Diffusion-GR2}, a recipe that converts our AR reasoning re-ranker (GR2) into a block-diffusion re-ranker. 
First, \emph{conversion fine-tuning} (CFT) adapts the AR-initialized diffusion model to denoise the answer into a valid permutation on its own, without an external constrained decoder. 
Next, \emph{on-policy distillation} (OPD) then supervises the model on its own decoded trajectories with dense per-token targets from the AR teacher. 
Finally, we apply a \emph{reinforcement-learning} (RL) stage against a re-ranking reward on top of OPD's on-policy policy. 
Experiments on Amazon Beauty demonstrate that Diffusion-GR2 recovers to near-parity with the AR re-ranker, while block-parallel decoding raises decode throughput by $2.4$--$3.5\times$ at the model's reasoning output length. Ablations show that CFT recovers most of the conversion gap, and that on-policy distillation further closes it to the AR reference.%
}

\date{June 30, 2026}
\correspondence{Mingfu Liang (\email{mingfuliang@meta.com}), Luke Simon (\email{lukesimon@meta.com}), Tianlong Chen (\email{tianlong@cs.unc.edu}), Xi Liu (\email{xliu1@meta.com})}

\begin{document}

\maketitle

\section{Introduction}
\label{section:intro}

Generative reasoning re-rankers~\citep{liang2026generative} built on large language models (LLMs)~\citep{radford2018improving,radford2019language} have recently been shown to refine the final stage of recommendation pipelines by reasoning explicitly over a short candidate list before committing to an ordering.
A reasoning re-ranker takes a user's purchase history together with a pre-ranked list of candidate items produced by a retriever, generates a chain-of-thought that justifies a re-ordering, and emits a permutation of the candidates.
The reasoning is what makes these models accurate: it grounds the decision in item semantics and user intent rather than a single learned score.
It is also what makes them expensive. An autoregressive (AR) decoder spends one sequential forward pass per reasoning token, and the reasoning trace is far longer than the ranking it ultimately produces. In a production re-ranking setting, where the same model is queried for every impression, this sequential cost is the binding constraint.

\begin{figure*}[t]
    \centering
    \includegraphics[width=0.98\textwidth]{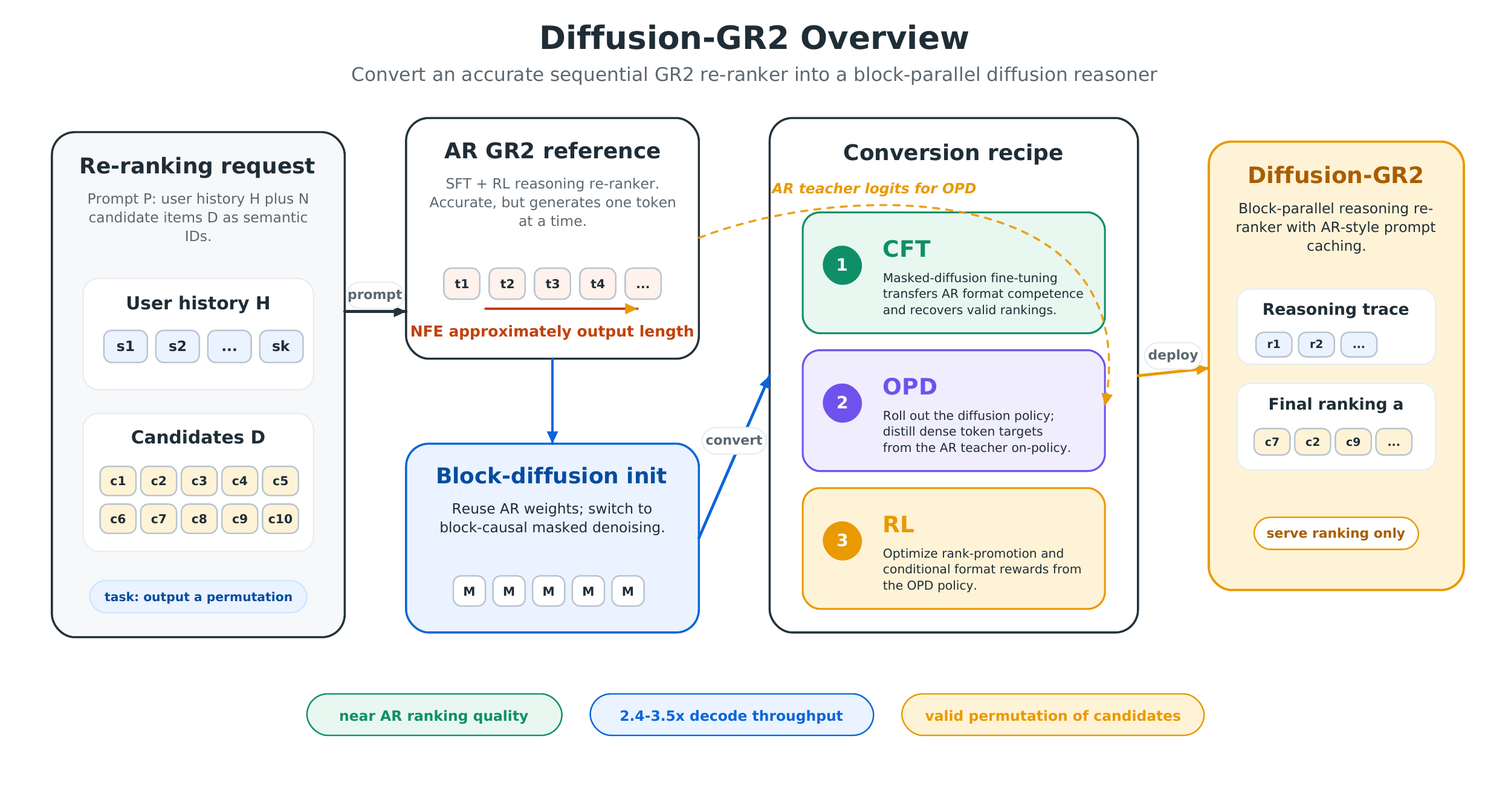}
    \caption{\textbf{Overview of Diffusion-GR2.} We start from the AR GR2 reasoning re-ranker, initialize a block-diffusion decoder from its weights, and recover the conversion gap through conversion fine-tuning (CFT), on-policy distillation (OPD), and reinforcement learning (RL). The final model preserves the GR2 re-ranking interface while reducing sequential decoding cost through block-parallel denoising.}
    \label{fig:overview}
\end{figure*}

\emph{Block-diffusion language models}~\citep{arriola2025interpolating,wu2025fast} suggest a way out.
Rather than walking the sequence left to right, they decode many positions at once and refine them over a small number of denoising steps, committing the most confident positions at each step.
This is attractive precisely where AR is most expensive, i.e., the long reasoning span, and it is the source of a large potential speedup.
But the speedup does not come for free. Replacing the AR decoder with a diffusion decoder \emph{drops ranking accuracy}, and the central question of this paper is how to convert the re-ranker to a diffusion decoder while giving up as little accuracy as possible.
We study this trade-off directly: we measure the accuracy cost of the conversion and develop a recipe that minimizes it. We refer to the resulting system as \textbf{Diffusion-GR2}, the block-diffusion counterpart of our autoregressive re-ranker GR2.

The starting point is our trained AR re-ranker.
Following GR2, it is obtained by supervised fine-tuning (SFT) on high-quality reasoning traces over semantic-ID-grounded items, followed by reinforcement learning (RL) against a re-ranking reward;
the resulting model sets the accuracy bar we try to recover after conversion.
We initialize a block-diffusion model from this AR model and ask it to reproduce the same reasoning-and-ranking behavior under parallel decoding.

The accuracy cost of the conversion has two sources, and Diffusion-GR2 addresses each in turn. The first source is \emph{structural} and concentrates at the answer.
The final ranking must be a permutation of exactly the candidates that were given.
An AR decoder enforces this implicitly: it commits to one item at a time and can mask the items it has already used.
A diffusion decoder predicts the answer positions in parallel and scores them independently, so nothing prevents it from repeating an item, omitting one, or emitting an identifier that was never a candidate.
In practice this happens often enough to cost real accuracy. We address this failure mode with \emph{conversion fine-tuning} (CFT): because the diffusion model is initialized from the AR re-ranker, which produces valid permutations for free, fine-tuning it on the re-ranking data transfers this competence, and the converted model learns to denoise the answer span into a valid permutation on its own without an external constrained decoder, recovering most of the accuracy lost at conversion.

The second source is \emph{distributional}.
After we fine-tune the converted model on fixed teacher trajectories, training is off-policy relative to inference: the model is trained on teacher reasoning traces and ground-truth answers, but at inference it decodes its \emph{own} trajectories under block-diffusion decoding. The mismatch leaves a residual gap.
We shrink it with \emph{on-policy distillation} (OPD): the converted model generates trajectories under its real block-diffusion decoding distribution, and the AR model supplies dense, per-token supervision on exactly those samples.
Because supervision is computed on the model's own outputs, training matches the distribution the model is evaluated under. Finally, a \emph{reinforcement-learning} stage against a re-ranking reward recovers the remaining margin;
consistent with our experience training the AR model, RL is most effective when it starts from OPD's healthy on-policy policy rather than cold.

We summarize our contributions as follows.
\begin{itemize}
    \item \textbf{Speed--accuracy trade-off.} We frame and quantify the cost of replacing the AR decoder of a reasoning re-ranker with a block-diffusion decoder, and present Diffusion-GR2, a recipe that minimizes the accuracy cost while retaining most of the speedup.
    \item \textbf{Conversion fine-tuning (CFT).} We adapt the AR-initialized diffusion model so that it emits valid permutations under parallel decoding without an external constrained decoder; by ablation, CFT accounts for most of the recovered accuracy.
    \item \textbf{On-policy distillation, then RL.} We show the residual off-policy gap is minimized by on-policy distillation (OPD), which recovers the bulk of the gap on-distribution, with a reinforcement-learning stage applied on top of OPD to recover the remaining margin.
    \item \textbf{Accuracy--latency frontier.} We validate Diffusion-GR2 on Amazon Beauty, showing that the converted model retains near-AR accuracy at $2.4$--$3.5\times$ higher decode throughput.
\end{itemize}

The remainder of this paper is organized as follows. \Cref{section:prelim} recaps the AR reasoning re-ranker and block-diffusion language models and fixes notation. \Cref{section:method} presents the Diffusion-GR2 conversion pipeline: the trade-off and its structural cause (\cref{subsec:tradeoff}), conversion fine-tuning (\cref{subsec:pcd}), on-policy distillation (\cref{subsec:opd}), and the reinforcement-learning stage (\cref{subsec:rl}). \Cref{section:exp} reports experiments and analyses on Beauty. \Cref{section:related} reviews related work, and \cref{section:conclusion} concludes. Additional algorithmic details, derivations, and hyperparameters are deferred to the Appendix.

\section{Preliminaries and Problem Setup}
\label{section:prelim}

\subsection{Reasoning Re-ranking}
\label{subsec:rerank}

We study reasoning-enabled re-ranking over a fixed candidate set, following the GR2 formulation. Each instance consists of a user purchase history $H = (s_{v_1}, \dots, s_{v_k})$ and a pre-ranked candidate list $D = (s_{y_1}, \dots, s_{y_N})$ of $N$ items produced by a retriever, where each item is represented by a semantic identifier (SID)~\citep{rajput2023recommender} grounded in an RQ-VAE tokenizer~\citep{lee2022autoregressive}. The goal is to re-order $D$ so that the ground-truth next item is promoted toward the top. A re-ranker is a conditional distribution $\pi_\theta(o \mid P(H, D))$ over an output $o = (\tau, a)$, where $\tau = (r_1, \dots, r_M)$ is a chain-of-thought reasoning trace and $a = (a_1, \dots, a_N)$ is the \emph{answer}: a permutation of the $N$ candidate identifiers. We use $N = 10$ throughout, consistent with the GR2 setup. The prompt $P(H, D)$ is rendered in a chat format with an expert system role, SID-grounded items with title and category metadata, and a structured output specification; we reuse the GR2 template verbatim and reproduce it in \cref{appendix:template}.

\subsection{The Autoregressive Re-ranker (GR2)}
\label{subsec:argr2}

Our starting model is the GR2 re-ranker, an LLM (Qwen3-8B~\citep{yang2025qwen3} unless stated otherwise) trained in two post-training stages. First, \emph{supervised fine-tuning} on high-quality reasoning traces generated by a larger teacher LLM and filtered by rejection sampling teaches the model to produce SID-grounded reasoning and a ranked list, with the language-modeling loss decoupled between reasoning and answer tokens. Second, \emph{reinforcement learning} with a re-ranking reward, i.e., a rank-promotion term combined with a conditional format reward, optimized with DAPO~\citep{yu2025dapo}, refines the policy to directly improve ranking quality. The SFT+RL model is the accuracy reference we aim to recover after conversion. The conversion to a block-diffusion model is weight-preserving: it copies the transformer tensors of this AR re-ranker one-to-one and only rewrites the configuration (block size, mask token, and the block-diffusion attention pattern), so that all adaptation to parallel decoding happens in the subsequent fine-tuning rather than at the repacking step. Crucially, the AR decoder enforces answer validity for free: at each step it emits a single identifier and can mask the identifiers it has already committed, so the answer is always a permutation of $D$.

\subsection{Block-Diffusion Language Models}
\label{subsec:bdlm}

A block-diffusion (or masked-diffusion) language model generates a sequence by iteratively denoising masked positions~\citep{austin2021structured,lou2024discrete,shi2024simplified,sahoo2024simple}. The sequence is partitioned into blocks~\citep{arriola2025block}; within a block, all masked positions are predicted in parallel from the current partially-denoised context, and a subset is \emph{committed} at each denoising step according to a confidence criterion (a per-position top-token probability threshold $\tau$), while the rest are re-masked and resolved in later steps (\cref{fig:inference-pdf}). Let $x^{(0)}$ be the fully-masked answer canvas and $x^{(s)}$ the state after $s$ denoising steps. At step $s$ the model produces position-wise distributions $p_\theta(\cdot \mid x^{(s)}, P)$ over the vocabulary $\mathcal{V}$ for every masked position, and a commitment rule $\mathcal{C}$ selects positions to fix:
\begin{equation}
\label{eq:diffstep}
x^{(s+1)} = \mathcal{C}\!\left(x^{(s)}, \, \{p_\theta(\cdot \mid x^{(s)}, P)\}\right).
\end{equation}
The number of sequential steps $S$ needed to fill a span of length $L$ can be far smaller than $L$, since many positions are committed per step; this is the source of the speedup over AR decoding, which requires exactly $L$ sequential passes. The cost is that positions committed in the same step are scored \emph{independently} given the shared context, which is harmless for free-form reasoning text but, as we show next, breaks the permutation structure of the answer.

\section{Method: The Diffusion-GR2 Conversion}
\label{section:method}

\subsection{Overview}
\label{subsec:overview}

\Cref{fig:overview} summarizes the Diffusion-GR2 pipeline. We begin from the trained AR GR2 re-ranker (\cref{subsec:argr2}) and initialize a block-diffusion model from its weights. We then close the conversion gap in three stages: \emph{conversion fine-tuning} (\cref{subsec:pcd}) adapts the model as a diffusion reasoner and, by transferring the AR model's format competence, recovers most of the lost accuracy while making answers valid; \emph{on-policy distillation} (\cref{subsec:opd}) removes the residual off-policy gap; and a \emph{reinforcement-learning} stage (\cref{subsec:rl}) recovers the remaining margin to near-parity. The output is a re-ranker that matches the AR model's accuracy at a fraction of its sequential decode cost.

\subsection{The Block-Diffusion Decoder and Inference Acceleration}
\label{subsec:blockwise}

\begin{figure*}[t]
    \centering
    \includegraphics[width=0.98\textwidth]{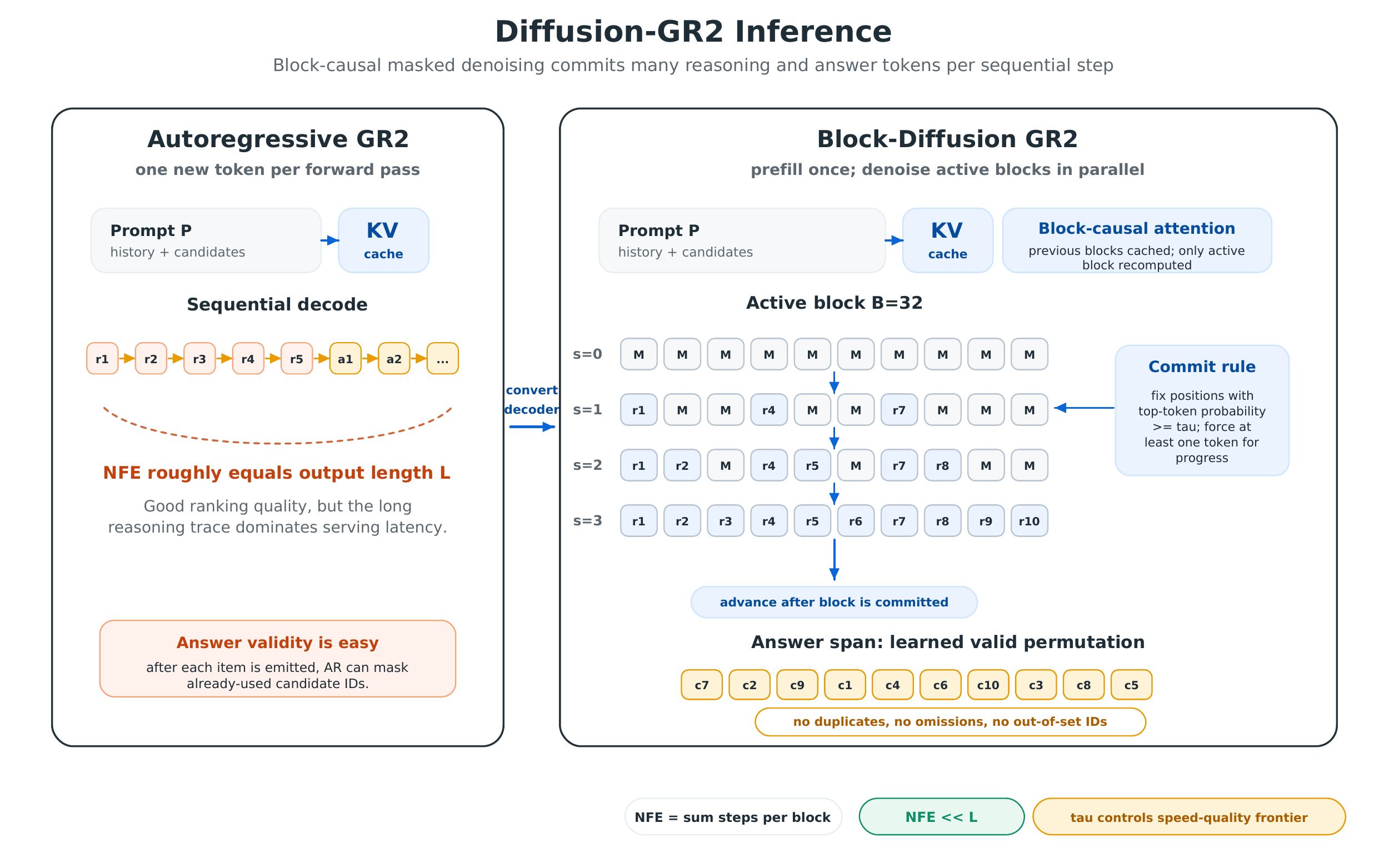}
    \caption{\textbf{Diffusion-GR2 inference.} AR GR2 emits one token per forward pass, whereas Diffusion-GR2 prefills the prompt once, reuses the KV cache for committed context, and denoises the active block in parallel. At each denoising step, high-confidence positions are committed according to threshold $\tau$, reducing the number of sequential forward passes from roughly the output length to the sum of denoising steps across blocks.}
    \label{fig:inference-pdf}
\end{figure*}

\Cref{fig:inference-pdf} illustrates the inference-time architecture of Diffusion-GR2, placing our block-diffusion decoder side by side with the autoregressive GR2 decoder it replaces; it is the reference for the design described throughout this subsection. To make the re-ranker fast, we replace its AR decoder with a \emph{block-diffusion} decoder while keeping the same backbone weights. In the style of Fast-dLLM~\citep{wu2025fast}, the decoder performs masked-diffusion denoising under attention that is \emph{causal across blocks}, rather than the fully bidirectional masked diffusion of LLaDA~\citep{zhu2026llada} or Dream~\citep{ye2025dream}. The choice is deliberate: the block structure is what turns the diffusion model's fewer-sequential-steps property into an actual wall-clock speedup at our long-prompt operating point. We discuss the structure and its consequences here; the accuracy cost of the conversion is taken up in \cref{subsec:tradeoff}.

\paragraph{Structure.} The response is partitioned into contiguous blocks of a fixed size $B$ (we use $B = 32$). Attention is \emph{bidirectional within a block} but \emph{causal across blocks}: a position in block $j$ attends to every position of its own block and to the committed (clean) context of all earlier blocks $j' < j$, but not to later blocks. Decoding proceeds block by block, and within each block the masked positions are denoised in parallel over a few steps (\cref{subsec:bdlm}). Once a block is fully committed the model advances to the next. Setting $B = 1$ recovers ordinary left-to-right autoregressive decoding as a special case; this degenerate setting, on the same weights, is exactly the AR baseline we compare against for speed (\cref{fig:inference-pdf}, left).

\paragraph{Why block diffusion: cacheable, prefill-amortized inference.} The across-block causal structure is what enables a key--value (KV) cache~\citep{wu2025fast}. Because a block never attends to later blocks, the keys and values of the prompt and of already-committed blocks are fixed for the remainder of decoding, so they can be cached once and reused, exactly as in autoregressive inference; each denoising step then recomputes attention only over the small active block against cached context, instead of over the full sequence. On top of this, confidence-thresholded parallel decoding commits every position whose top-token probability exceeds a threshold $\tau$ at each step (plus a forced arg-max to guarantee progress), so a block of $B$ positions is typically resolved in far fewer than $B$ forward passes. The number of sequential forward passes (NFE) to produce a response of length $L$ is therefore $\sum_{\text{blocks}} (\text{steps per block}) \ll L$, and $\tau$ tunes the trade-off: a higher $\tau$ commits fewer, more confident tokens per step (closer to AR, higher quality), a lower $\tau$ commits more per step (faster, until parsing degrades).

\paragraph{Contrast with fully bidirectional diffusion.} Decoder-only masked-diffusion language models with fully bidirectional attention (e.g.\ LLaDA~\citep{zhu2026llada}, Dream~\citep{ye2025dream}) attend over the entire sequence at every denoising step, including the prompt. This is marginally more flexible, i.e., any position may condition on any other at any step. However, it forecloses both accelerations above: there is no causal direction to cache against, so the KV cache cannot be reused across steps, and the long prompt must be re-encoded at every step. At our operating point this is decisive: the re-ranking prompt averages $\sim$2{,}200 tokens against a short, constrained answer, so per-step prompt re-encoding is precisely the cost we cannot afford. Block diffusion keeps the prompt and completed blocks cached and pays full attention only on the active block, recovering AR-style prefill amortization while still committing many answer tokens in parallel. We therefore adopt a block-diffusion decoder for Diffusion-GR2, accepting its slightly more constrained attention in exchange for inference that a fully bidirectional model cannot match at long-prompt, short-answer re-ranking.

\subsection{The Conversion Trade-off and Why Naive Decoding Loses Accuracy}
\label{subsec:tradeoff}

Converting the re-ranker to a diffusion decoder is a speed--accuracy trade-off, and the accuracy cost concentrates at the answer. The reasoning span and the answer span behave very differently under parallel decoding. Reasoning is free-form text: committing several tokens in the same step, conditioned on the shared context, is benign. The answer is not free-form: it is a permutation of a known set, and its positions are highly coupled. Because diffusion commits answer positions in parallel and scores them independently (\cref{eq:diffstep}), the decoder has no mechanism to enforce that the committed identifiers form a permutation of $D$. Three failure modes result:
\begin{itemize}
    \item \emph{Duplicates}: the same candidate identifier is committed at two ranks.
    \item \emph{Omissions}: a candidate never appears in the answer.
    \item \emph{Out-of-set identifiers}: a token that is a syntactically valid identifier, but not among the $N$ candidates of this query, is committed.
\end{itemize}
An AR decoder cannot make these mistakes, because it places one item at a time and masks what it has already used. This asymmetry is the first and largest part of the conversion gap: a naively converted model frequently produces malformed rankings, and the resulting accuracy falls well below the AR reference. The next section shows how the conversion itself recovers validity.

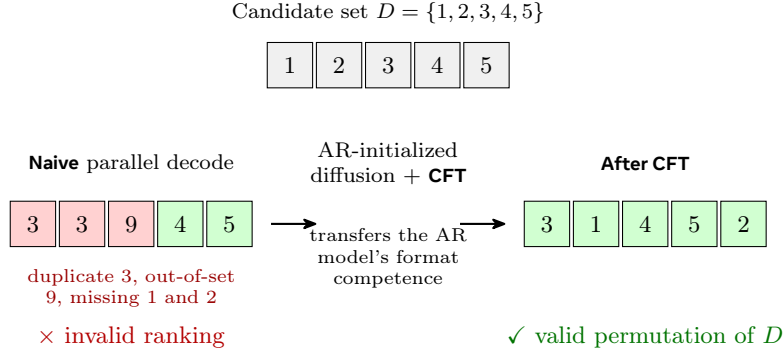
\begin{figure}[t]
    \centering
    \begin{tikzpicture}[font=\small, >={Stealth[round]},
      chip/.style={draw, minimum size=0.6cm, inner sep=0pt, font=\small},
      ok/.style={chip, fill=green!18},
      bad/.style={chip, fill=red!18},
    ]
    \node[font=\footnotesize] at (0,1.3) {Candidate set $D=\{1,2,3,4,5\}$};
    \foreach \v [count=\i from 0] in {1,2,3,4,5} {\node[chip, fill=gray!12] at (\i*0.65-1.3,0.6) {\v};}
    \node[font=\footnotesize, align=center] at (-3.4,-0.7) {\textbf{Naive} parallel decode};
    \node[bad] at (-4.7,-1.5) {3}; \node[bad] at (-4.05,-1.5) {3};
    \node[bad] at (-3.4,-1.5) {9}; \node[ok] at (-2.75,-1.5) {4}; \node[ok] at (-2.1,-1.5) {5};
    \node[font=\scriptsize, red!60!black, align=center, text width=3.2cm] at (-3.4,-2.35)
       {duplicate 3, out-of-set 9, missing 1 and 2};
    \node[font=\small, red!70!black] at (-3.4,-3.0) {$\times$~invalid ranking};
    \node[align=center, font=\footnotesize, text width=2.2cm] at (0,-0.7)
       {AR-initialized diffusion $+$ \textbf{CFT}};
    \node[align=center, font=\scriptsize, text width=2.4cm] at (0,-1.95)
       {transfers the AR model's format competence};
    \draw[->, thick] (-1.55,-1.5) -- (-0.95,-1.5);
    \draw[->, thick] (0.95,-1.5) -- (1.55,-1.5);
    \node[font=\footnotesize, align=center] at (3.4,-0.7) {\textbf{After CFT}};
    \node[ok] at (2.1,-1.5) {3}; \node[ok] at (2.75,-1.5) {1};
    \node[ok] at (3.4,-1.5) {4}; \node[ok] at (4.05,-1.5) {5}; \node[ok] at (4.7,-1.5) {2};
    \node[font=\small, green!45!black] at (3.4,-3.0) {$\checkmark$~valid permutation of $D$};
    \end{tikzpicture}
    \caption{\textbf{Stage 1: conversion fine-tuning (CFT).} Decoding the converted model \emph{naively} scores answer positions independently, so it routinely emits invalid rankings (duplicated, out-of-set, or missing identifiers). Because the diffusion model is initialized from the AR re-ranker, CFT transfers its format competence: the model learns to denoise the answer span into a valid permutation of the candidate set $D$ on its own, with no external constrained decoder.}
    \label{fig:cft}
\end{figure}

\subsection{Stage 1: Conversion Fine-Tuning (CFT)}
\label{subsec:pcd}

The failure modes of \cref{subsec:tradeoff} are not specific to re-ranking: diffusion language models are broadly unreliable at structured output, because they commit positions in parallel and score them independently rather than tracking what has already been emitted. A natural fix is to attach a \emph{constrained decoder} at inference that masks the answer logits to the candidate set; this guarantees syntactic validity, but it is external machinery bolted onto a model whose generative behavior is unchanged, and it does nothing for the \emph{order} the model was trained to produce.

We take a different route that is native to our setting: we recover validity from the conversion itself. Our diffusion re-ranker is not trained from scratch. It is initialized from the AR GR2 model, which produces valid permutations of $D$ for free through left-to-right masking (\cref{subsec:argr2}), and is then fine-tuned on the re-ranking data with a masked-diffusion objective over the assistant message, decoupling the loss between reasoning and answer tokens as in GR2. AR initialization transfers the source model's capabilities into the diffusion decoder, and we find that emitting a well-formed ranking is one of the behaviors it transfers: after conversion fine-tuning, the model learns to denoise the answer span into a valid permutation on its own, without a task-specific constrained decoder, recovering most of the accuracy lost at conversion.

The model is then free to spend its capacity on getting the \emph{order} right rather than on remembering to produce a well-formed list. \Cref{fig:cft} contrasts a naively decoded answer, i.e., with duplicated, out-of-set, and missing identifiers, against the valid permutation the model emits after CFT. A residual gap to the AR reference nonetheless remains, which we turn to next.

\begin{figure}[t]
    \centering
    \begin{tikzpicture}[font=\small, >={Stealth[round]},
      cell/.style={draw, minimum size=0.6cm, inner sep=0pt},
      tok/.style={cell, fill=dgr2blue!30},
      msk/.style={cell, fill=gray!20},
      bx/.style={draw, rounded corners, align=center, inner sep=6pt, fill=gray!10},
    ]
    \node[bx, text width=5.4cm] (T) at (2.0,2.7) {\textbf{AR teacher} $\pi^{\mathrm{AR}}_\phi$ (frozen)\\ causal, \emph{clean} context};
    \foreach \i [count=\c from 0] in {1,2,3,4,5} {\node[tok] at (\c*1.0,1.4) {$o_{\i}$};}
    \node[font=\scriptsize, align=center] at (-1.2,1.4) {clean\\ response};
    \node[bx, text width=5.4cm] (S) at (2.0,-2.7) {\textbf{Student} $\pi_\theta$ (block-diffusion)\\ \emph{masked} canvas};
    \node[tok] at (0,-1.4) {$o_1$};
    \node[msk] at (1,-1.4) {M};
    \node[tok] at (2,-1.4) {$o_3$};
    \node[msk] at (3,-1.4) {M};
    \node[msk] at (4,-1.4) {M};
    \node[font=\scriptsize, align=center] at (-1.2,-1.4) {masked\\ canvas};
    \draw[->, thick] (T.south) -- (2.0,1.75) node[midway, right, font=\scriptsize]{score};
    \draw[->, thick] (S.north) -- (2.0,-1.75) node[midway, right, font=\scriptsize]{decode};
    \foreach \c in {1,3,4}{
      \draw[<->, dashed] (\c,1.05) -- (\c,-1.05)
        node[midway, fill=white, inner sep=1pt, font=\scriptsize] {KL};
    }
    \node[bx, fill=white, text width=3.7cm, align=center, font=\footnotesize] at (7.5,0)
      {\textbf{OPD objective}\\[3pt]
       $\displaystyle \min_\theta \mathrm{KL}\!\big(\pi^{\mathrm{AR}}_\phi \,\big\|\, \pi_\theta\big)$\\[3pt]
       forward KL \emph{only at masked positions}; both logits right-shifted by one ($t\!\to\!$ token $t$)};
    \end{tikzpicture}
    \caption{\textbf{Stage 2: on-policy distillation (OPD).} The AR teacher is scored on the \emph{clean} response under causal attention, while the student is scored on the \emph{masked} block-diffusion canvas it decodes. The forward KL $\mathrm{KL}(\pi^{\mathrm{AR}}_\phi\,\|\,\pi_\theta)$ is applied only at the student's masked positions, with both logits right-shifted by one so position $t$ scores token $t$, so a single AR teacher supervises the diffusion student token-for-token and removes the off-policy gap left by CFT.}
    \label{fig:opd}
\end{figure}
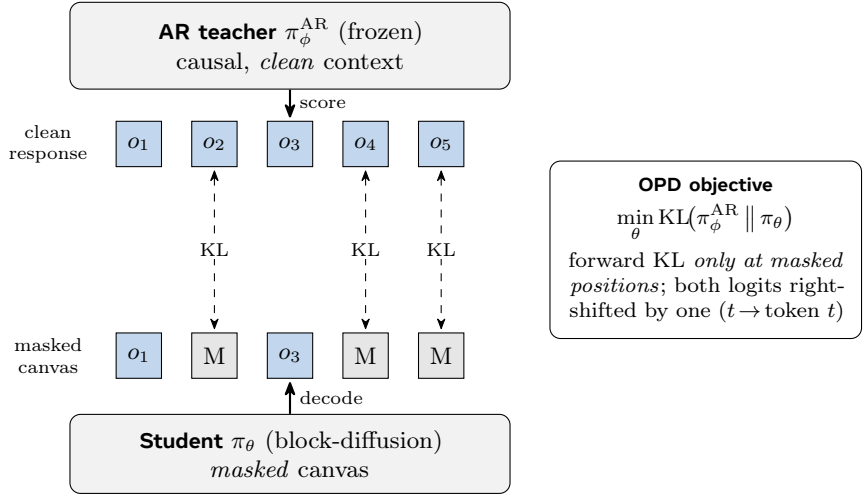

\subsection{Stage 2: On-Policy Distillation (OPD)}
\label{subsec:opd}

With answer validity recovered, a residual gap to the AR reference remains, for a structural reason: conversion fine-tuning is \emph{off-policy}. 
The model is trained on fixed trajectories (teacher reasoning traces and ground-truth answers), but at inference it decodes its \emph{own} trajectories, i.e., the denoising order, the partially-decoded contexts it conditions on, and its own commitments are not the contexts it saw during fine-tuning, and this train/inference mismatch shows up as lost ranking accuracy. 
This is precisely the \emph{exposure bias} of sequence models trained by teacher forcing~\citep{ranzato2016sequence}: supervising on fixed teacher trajectories is a form of sequence-level knowledge distillation~\citep{kim2016sequence} that never exposes the student to the partially-decoded, self-generated contexts it must condition on at inference, so small per-step deviations compound over the trajectory.
The standard remedy is to make training \emph{on-policy}, i.e., to supervise the student on its own rollouts rather than the teacher's, as in DAgger~\citep{ross2011reduction} for imitation learning and generalized knowledge distillation~\citep{agarwal2024policy} for language models. 
We follow this route with on-policy distillation~\citep{lu2025onpolicydistillation}. The converted model generates trajectories under its \emph{real} decoding distribution (block-diffusion denoising as used at inference) and the AR model supplies dense, per-token supervision on exactly those samples. 

Aligning the two models is not immediate: the AR teacher and the block-diffusion student use different attention and masking---the teacher factorizes left-to-right over a \emph{clean} sequence, whereas the student predicts \emph{masked} positions from a partially-denoised, block-bidirectional canvas. 
We resolve this mismatch in three steps inspired by \citet{su2026data}: (i)~the teacher is evaluated on the clean, mask-free response under its native causal attention, while the student is evaluated on the masked canvas it actually decodes; 
(ii)~because both models' logits are AR-derived (position $t$ scores token $t{+}1$), we right-shift both by one so that the distribution at position $t$ scores token $t$; 
and (iii)~the forward KL is applied \emph{only at the student's masked response positions}, where the teacher's clean-context distribution supplies the per-token target. 
This lets a single AR teacher supervise the diffusion student token-for-token despite their different decoding masks.
Concretely, let $o \sim \pi_\theta(\cdot \mid P)$ be a trajectory sampled on-policy from the converted model, with denoising states $\{x^{(s)}\}$. For each committed position $t$ we minimize the token-level divergence between the converted model and the AR teacher $\pi_\phi^{\mathrm{AR}}$ evaluated on the on-policy context:
\begin{equation}
\label{eq:opd}
\mathcal{L}_{\mathrm{OPD}}(\theta) \;=\; \mathbb{E}_{P \sim \mathcal{D}}\;\mathbb{E}_{o \sim \pi_\theta(\cdot \mid P)} \left[\; \frac{1}{|o|}\sum_{t=1}^{|o|} \mathrm{KL}\!\left(\pi_\phi^{\mathrm{AR}}(\cdot \mid P, o_{<t}) \,\big\|\, \pi_\theta(\cdot \mid P, o_{<t})\right)\right].
\end{equation}
Because the supervision is computed on the model's own decoded outputs, training matches the distribution the model is actually evaluated under, directly attacking the off-policy mismatch described above. Dense per-token targets also give the optimizer far more signal than a single ranking scalar. The teacher is the SFT+RL AR re-ranker, so OPD pulls the converted model back toward that model's accuracy. \Cref{fig:opd} illustrates this on-policy distillation loop. OPD recovers the bulk of the residual gap (\cref{subsec:main-results}).

\subsection{Stage 3: Reinforcement Learning}
\label{subsec:rl}

A final reinforcement-learning stage optimizes the re-ranking objective directly and recovers the last of the margin. This instantiates, for re-ranking, the reinforcement-learning-from-verifiable-rewards (RLVR) paradigm behind recent reasoning LLMs~\citep{guo2025deepseek}: unlike RLHF against a learned preference model~\citep{ouyang2022training}, our reward is a \emph{checkable} ranking signal. We reuse the GR2 reward: a rank-promotion reward that measures how much the ground-truth target $s_{v_{n+1}}$ is lifted by re-ranking,
\begin{equation}
\label{eq:rankreward}
R_{\mathrm{rank}} \;=\; \frac{r^{D}_{s_{v_{n+1}}} - r^{o}_{s_{v_{n+1}}}}{N},
\end{equation}
where $r^{D}$ and $r^{o}$ are the ranks of the target in the pre-ranked and re-ranked lists, combined with a conditional format reward $R_{\mathrm{fmt}}$ that is granted only when re-ranking strictly improves the target's rank, or when an already-top-1 target is preserved at rank~1, so as not to reward order-preserving degeneracy.

\begin{figure}[t]
    \centering
    \begin{tikzpicture}[font=\small, >={Stealth[round]},
      bx/.style={draw, rounded corners, align=center, inner sep=6pt},
      slot/.style={draw, minimum width=0.5cm, minimum height=0.42cm, inner sep=0pt},
      tgt/.style={slot, fill=dgr2blue!45},
      oth/.style={slot, fill=gray!12},
    ]
    \node[bx, fill=gray!10, text width=1.9cm] (pol) at (0,0) {\textbf{Policy} $\pi_\theta$};
    \foreach \p in {0,1,2,3,4}{\node[oth] at (2.65+\p*0.52,0.9){};}
    \node[tgt] at (2.65,0.9){}; \node[font=\scriptsize, left] at (2.3,0.9){$o_1$};
    \foreach \p in {0,1,2,3,4}{\node[oth] at (2.65+\p*0.52,0.0){};}
    \node[tgt] at (2.65,0.0){}; \node[font=\scriptsize, left] at (2.3,0.0){$o_2$};
    \foreach \p in {0,1,2,3,4}{\node[oth] at (2.65+\p*0.52,-0.9){};}
    \node[tgt] at (2.65+3*0.52,-0.9){}; \node[font=\scriptsize, left] at (2.3,-0.9){$o_3$};
    \node[font=\scriptsize, align=center, text width=4.2cm] at (3.7,-1.75)
       {$G$ rollouts; target (blue) lifted toward rank~1 $\Rightarrow$ higher $R_{\mathrm{rank}}$};
    \draw[->, thick] (pol.east) -- (1.85,0);
    \node[bx, fill=orange!10, text width=3.5cm] (rew) at (7.9,1.7)
      {$R=R_{\mathrm{rank}}+\alpha R_{\mathrm{fmt}}$,\quad $R_{\mathrm{rank}}=\dfrac{r^{D}-r^{o}}{N}$};
    \node[bx, fill=green!8, text width=3.5cm] (adv) at (7.9,-0.4)
      {group-normalized advantage\\ $\hat A_i=(R_i-\mathrm{mean})/\mathrm{std}$};
    \node[bx, fill=blue!8, text width=3.5cm] (ppo) at (7.9,-2.4)
      {TraceRL: $\rho_{i,t}$ by \emph{replaying} the denoising trace; PPO update};
    \draw[->, thick] (5.2,0) -- (rew.west);
    \draw[->, thick] (rew) -- (adv);
    \draw[->, thick] (adv) -- (ppo);
    \end{tikzpicture}
    \caption{\textbf{Stage 3: reinforcement learning.} For each prompt the policy samples a group of $G$ rollouts; the rank-promotion reward $R_{\mathrm{rank}}=(r^{D}-r^{o})/N$ (plus a conditional format reward) scores how far the target is lifted toward rank~1. Advantages are group-normalized, and the importance ratio $\rho_{i,t}$ is computed by replaying each rollout's recorded denoising trace before the PPO update. Applied on top of OPD, this recovers the last margin to the AR teacher.}
    \label{fig:rl}
\end{figure}
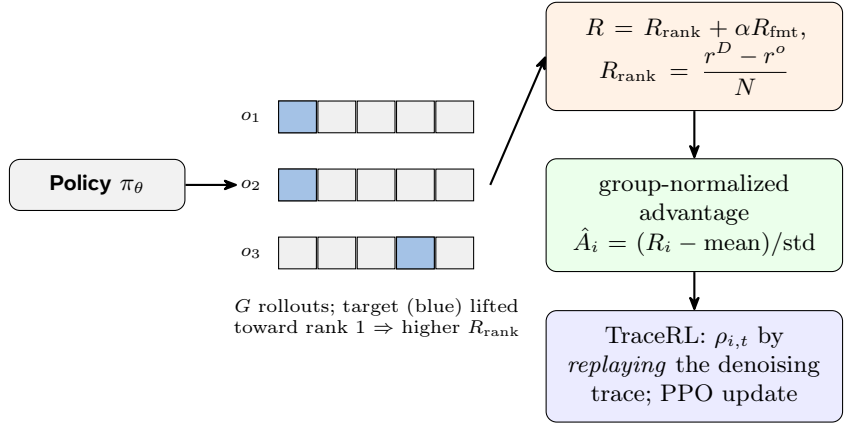

Inspired by TraceRL~\citep{wang2025revolutionizing}, we optimize the diffusion policy with a trajectory-level GRPO/DAPO-style objective~\citep{schulman2017proximal,shao2024deepseekmath,yu2025dapo} over groups of $G$ sampled trajectories.
Rather than scoring each trajectory with a single mean-field forward pass, we evaluate the importance ratio $\rho_{i,t}$ by \emph{replaying the recorded denoising trace} of each rollout: for each committed position $t$ we reconstruct the masked canvas at the denoising step at which $t$ was committed and score it under the model's inference-time (block-causal) attention and autoregressive token-shift.
Because the converted model emits valid permutations, $R_{\mathrm{rank}}$ is well-defined on essentially every rollout;
an invalid or overflowing rollout instead receives a reward of $-1$ and is kept in its group, and we drop only groups with zero reward variance, for which the normalized advantage is undefined:
\begin{equation}
\label{eq:grpo}
\mathcal{J}(\theta) = \mathbb{E}_{P \sim \mathcal{D},\, \{o_i\}_{i=1}^{G} \sim \pi_{\theta_{\mathrm{old}}}(\cdot \mid P)} \left[\frac{1}{\sum_i |o_i|}\sum_{i=1}^{G}\sum_{t=1}^{|o_i|} \min\!\big(\rho_{i,t}(\theta)\hat{A}_{i,t},\; \mathrm{clip}(\rho_{i,t}(\theta), 1-\varepsilon, 1+\varepsilon)\hat{A}_{i,t}\big)\right],
\end{equation}
where $\rho_{i,t}(\theta)$ is the trajectory-replay importance ratio defined above and $\hat{A}_{i,t} = (R_i - \mathrm{mean}\{R_j\})/\mathrm{std}\{R_j\}$ is the group-normalized advantage. \Cref{fig:rl} summarizes the stage.

The RL stage is most effective applied \emph{on top of} OPD. OPD provides a healthy, on-distribution policy with non-degenerate trajectories, and from that starting point RL improves the ranking metric and closes more of the residual gap than OPD alone. The headline gap-minimizing recipe is therefore \textbf{OPD\,$\rightarrow$\,RL}: dense on-policy distillation to recover the bulk, then RL to optimize the metric directly for the remaining margin.

\section{Experiments}
\label{section:exp}

We organize the evaluation around three questions:
\begin{itemize}
    \item \textbf{Q1}: How much ranking accuracy does converting the AR re-ranker to a block-diffusion decoder cost, and how much do conversion fine-tuning and on-policy distillation recover?
    \item \textbf{Q2}: What inference speedup does the block-diffusion decoder buy over autoregressive decoding, and at what accuracy?
    \item \textbf{Q3}: Does the converted model reason faithfully, or reach its rankings with degraded filler?
\end{itemize}

\subsection{Experimental Setup}
\label{subsec:setup}

\paragraph{Datasets.}
We evaluate on the Amazon Review \emph{Beauty} dataset~\citep{mcauley2015image} under the standard TIGER protocol~\citep{rajput2023recommender}: 5-core filtering, chronological ordering, and a leave-one-out train/validation/test split (\cref{table:datastats}). Each item is encoded as a 4-token RQ-VAE semantic identifier. For each user, a retriever returns a top-10 candidate list, which the re-ranker reorders to promote the ground-truth next item; the retriever's ordering is the \emph{pre-rank floor}. The test set contains $n = 1{,}615$ users.

\begin{table}[t]
    \centering
    \begin{NiceTabular}{lccc}
    \toprule
    Dataset & \#Users & \#Items & Avg.\ Seq.\ Len. \\
    \midrule
    Beauty & 22{,}363 & 12{,}101 & 8.87 \\
    \bottomrule
    \end{NiceTabular}
    \caption{Statistics of the Amazon Review Beauty dataset used for re-ranking.}
    \label{table:datastats}
\end{table}

\paragraph{Metrics.}
We report Recall@$K$ ($K \in \{1, 3\}$) and NDCG@$3$ on the re-ranked list, computed with a single relevant item (IDCG $=1$) and a log-base-2 discount, consistent with GR2. For efficiency we report decode throughput (tokens per second) at the model's reasoning output length ($\sim$130 tokens), on a single H100-80G GPU with \texttt{torch.compile}.

\paragraph{Systems.}
All systems share the Qwen3-8B backbone. The reference points are the \emph{pre-rank floor} (the retriever's order, no re-ranking) and the \emph{AR GR2 re-ranker} (the autoregressive teacher, our accuracy target). The converted models are: the block-diffusion model decoded \emph{without} conversion fine-tuning (\emph{naive}); and Diffusion-GR2 after each stage, i.e., conversion fine-tuning ($+$CFT), on-policy distillation ($+$OPD), and reinforcement learning ($+$OPD\,$\rightarrow$\,RL). Unless noted, accuracy is measured under quality-preserving decoding (confidence threshold $1.0$); the speed--accuracy trade-off of parallel decoding is studied in \cref{subsec:latency}.

\subsection{Q1: Recovering the Conversion Accuracy Gap}
\label{subsec:main-results}

\Cref{table:main-beauty} reports re-ranking accuracy on Beauty; read top to bottom, it traces the conversion. The pre-rank floor is Recall@1 $0.2811$, and the AR GR2 re-ranker reaches $0.2960$, i.e., the accuracy we aim to recover. Decoding the converted model \emph{naively}, in parallel and without conversion fine-tuning, emits a valid ranking for only $\sim$0.1\% of queries (valid-JSON rate $0.001$) and never reorders; the malformed outputs fall back to the retriever's order, so naive decoding collapses exactly to the pre-rank floor ($0.2811$ Recall@1) and forfeits the entire re-ranking gain (\cref{subsec:tradeoff}). Conversion fine-tuning recovers the great majority of the gap: Diffusion-GR2 $+$CFT reaches Recall@1 $0.2930$, within $0.0030$ of the AR reference. On-policy distillation closes most of what remains: $+$OPD reaches $0.2944$, within $0.0016$ of the AR teacher. A reinforcement-learning stage on top of OPD (OPD\,$\rightarrow$\,RL) recovers the final margin, reaching Recall@1 $0.2951$ and essentially matching the AR teacher. The deeper ranking metrics tell an even stronger story: on Recall@3 and NDCG@3 the converted model already matches the AR teacher after conversion fine-tuning (both $0.5651$ and $0.4497$) and then \emph{surpasses} it in the later stages: $+$OPD lifts Recall@3 to $0.5658$, and OPD\,$\rightarrow$\,RL attains the best Recall@3 ($0.5671$) and NDCG@3 ($0.4517$) in the table, above the AR teacher's $0.5651$ and $0.4497$. In other words, the recipe does not merely recover top-$1$ accuracy to near-parity; on the broader top-$3$ ranking quality it modestly \emph{exceeds} the teacher, as reinforcement learning redistributes probability among the top candidates rather than only the single top slot. Two observations close the question. First, it is conversion fine-tuning, not an external constrained decoder, that recovers answer validity and the bulk of the accuracy; on-policy distillation then supplies dense per-token targets on the model's own decoded trajectories, directly correcting the off-policy mismatch of \cref{subsec:opd} and lifting accuracy nearest to the teacher. Second, the band is narrow: the AR teacher sits only $0.0149$ Recall@1 above the pre-rank floor, so the converted model operates in a near-saturated regime in which OPD already reaches near-parity. This answers \textbf{Q1}.

\begin{table*}[t]
    \centering
    \begin{NiceTabular}{lccc}
    \CodeBefore
    \rectanglecolor{metabg}{7-1}{7-4}
    \Body
    \toprule
    Method & Recall@1 & Recall@3 & NDCG@3 \\
    \midrule
    Pre-rank floor (retriever)                     & $\nm{0.2811}$ & $\nm{0.5591}$ & $\nm{0.4401}$ \\
    AR GR2 (reference)                             & $\bm{0.2960}$ & $\nm{0.5651}$ & $\nm{0.4497}$ \\
    \midrule
    Diffusion-GR2, naive (no CFT)                  & $\nm{0.2811}$ & $\nm{0.5591}$ & $\nm{0.4401}$ \\
    Diffusion-GR2 $+$CFT                           & $\nm{0.2930}$ & $\nm{0.5651}$ & $\nm{0.4497}$ \\
    Diffusion-GR2 $+$CFT $+$OPD                    & $\nm{0.2944}$ & $\nm{0.5658}$ & $\nm{0.4497}$ \\
    Diffusion-GR2 $+$CFT $+$OPD\,$\rightarrow$\,RL & $\nm{0.2951}$ & $\bm{0.5671}$ & $\bm{0.4517}$ \\
    \bottomrule
    \end{NiceTabular}
    \caption{Re-ranking accuracy on \textbf{Amazon Beauty} (Qwen3-8B backbone; quality-preserving decoding). The AR GR2 teacher (bold) is the accuracy target; the pre-rank floor is the retriever's order. Conversion fine-tuning recovers most of the conversion gap, and OPD\,$\rightarrow$\,RL (highlighted) reaches Recall@1 $0.2951$, essentially matching the teacher.}
    \label{table:main-beauty}
\end{table*}

\subsection{Q2: The Accuracy--Latency Frontier}
\label{subsec:latency}

\begin{figure}[t]
    \centering
    \includegraphics[width=\columnwidth]{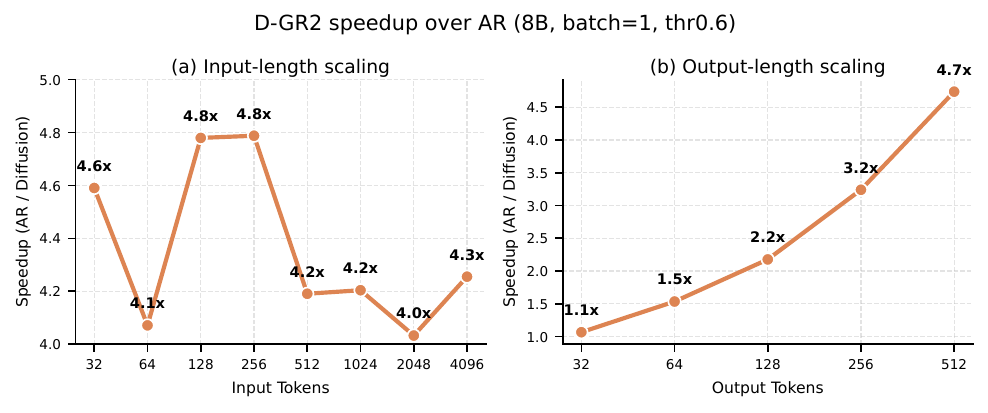}
    \caption{Decode speedup of Diffusion-GR2 over the AR re-ranker (Qwen3-8B, single request, $\tau{=}0.6$, \texttt{torch.compile}), sweeping (a) input length (output fixed at $512$) and (b) output length (input fixed at $\sim$2{,}400). The diffusion advantage grows with output length, where AR pays one sequential pass per token, and stays roughly flat in input length, reflecting that the gain comes from parallel \emph{decoding} rather than prefill.}
    \label{fig:speedup-scaling}
\end{figure}

\Cref{table:latency} compares the AR re-ranker against Diffusion-GR2 on decode throughput at the reasoning output length ($\sim$130 tokens). The AR re-ranker decodes sequentially at $71$ tokens per second. Diffusion-GR2 decodes its answer in parallel and reaches $172$--$246$ tok/s, i.e., a $2.4$--$3.5\times$ speedup, by committing multiple tokens per forward pass. The confidence threshold $\tau$ governs the trade-off: at $\tau{=}0.9$ the valid-JSON rate stays at $1.0$ and Recall@1 is $0.2950$ at $2.4\times$; at $\tau{=}0.6$ throughput rises to $3.3\times$ (Recall@1 $0.2942$); below $\tau{\le}0.4$ parsing degrades and accuracy drops. Diffusion-GR2 thus delivers near-AR ranking accuracy at $2.4$--$3.5\times$ higher throughput; combined with the quality-preserving operating point of \cref{table:main-beauty} (Recall@1 $0.2951$ at AR-level latency), this traces the accuracy--latency frontier. \Cref{fig:speedup-scaling} breaks this speedup down by sequence length: the diffusion advantage \emph{grows with the output length}---where AR pays one sequential forward pass per token---while staying roughly flat across input length, confirming that the gain comes from parallel \emph{decoding} rather than prefill. This scaling favors the long reasoning traces that dominate re-ranking latency, and points to even larger gains at higher reasoning budgets. This answers \textbf{Q2}.

\begin{table}[t]
    \centering
    \begin{NiceTabular}{lccc}
    \CodeBefore
    \rectanglecolor{metabg}{3-1}{3-4}
    \Body
    \toprule
    Decoding & Recall@1 & Throughput (tok/s) & Speedup \\
    \midrule
    AR GR2                       & $\nm{0.2960}$ & $\nm{71}$  & $1.0\times$ \\
    Diffusion-GR2 ($\tau{=}0.9$) & $\nm{0.2950}$ & $\nm{172}$ & $2.4\times$ \\
    Diffusion-GR2 ($\tau{=}0.6$) & $\nm{0.2942}$ & $\nm{234}$ & $3.3\times$ \\
    Diffusion-GR2 ($\tau{=}0.4$) & $\nm{0.2936}$ & $\nm{246}$ & $3.5\times$ \\
    \bottomrule
    \end{NiceTabular}
    \caption{Accuracy--latency frontier on Beauty (Qwen3-8B, $n{=}1615$, single H100-80G, \texttt{torch.compile}, $\sim$2{,}215-token prompt, $\sim$130-token reasoning output). The AR re-ranker decodes sequentially; Diffusion-GR2 decodes in parallel with block size $32$ and confidence threshold $\tau$. Diffusion-GR2 is $2.4$--$3.5\times$ faster; $\tau{=}0.9$ (highlighted) keeps the valid-JSON rate at $1.0$, while $\tau{\le}0.4$ degrades parsing.}
    \label{table:latency}
\end{table}

\subsection{Q3: Reasoning Quality}
\label{subsec:reasoning-quality}

Because free-form reasoning is the hard case for parallel decoding, we ask whether Diffusion-GR2 reaches its rankings with genuine reasoning or with degraded filler. We run an LLM-as-judge study~\citep{zheng2023judging} over $50$ paired, blind comparisons of AR and Diffusion-GR2 reasoning traces on Beauty, scoring three $1$--$5$ quality axes, i.e., history grounding, internal consistency, and logical flow, plus a binary identifier-correctness check ($0$/$1$, where $1$ denotes no history/candidate-identifier confusion). \Cref{table:reasoning} reports the per-axis means. Diffusion-GR2 tracks the AR re-ranker closely on every axis: history grounding ($4.34$ vs.\ $4.50$), internal consistency ($4.94$ vs.\ $4.94$), logical flow ($4.40$ vs.\ $4.54$), and identifier correctness ($1.00$ vs.\ $1.00$). A blind pairwise preference over the same $50$ pairs is AR $17$ / Diffusion-GR2 $9$ / tie $24$, with a $100\%$ parse rate and no history/candidate-identifier confusion, i.e., no systematic degradation in reasoning quality. This answers \textbf{Q3}.
\begin{table}[t]
    \centering
    \begin{NiceTabular}{lcc}
    \toprule
    Reasoning axis & AR GR2 & Diffusion-GR2 \\
    \midrule
    History grounding ($1$--$5$)      & $4.50$ & $4.34$ \\
    Internal consistency ($1$--$5$)   & $4.94$ & $4.94$ \\
    Logical flow ($1$--$5$)           & $4.54$ & $4.40$ \\
    Identifier correctness ($0$/$1$)  & $1.00$ & $1.00$ \\
    \bottomrule
    \end{NiceTabular}
    \caption{Reasoning-quality comparison on Beauty (LLM-as-judge, $50$ paired blind evaluations). The first three axes are scored $1$--$5$; identifier correctness is a binary $0$/$1$ check (fraction of traces with no identifier confusion). Diffusion-GR2 tracks the AR re-ranker on every axis; the blind pairwise preference is AR $17$ / Diffusion-GR2 $9$ / tie $24$, indicating no systematic degradation.}
    \label{table:reasoning}
\end{table}

\section{Related Work}
\label{section:related}

\subsection{Generative and Reasoning Re-rankers}
Generative recommendation~\citep{zhou2025efficiency,zhang2025reasonrec, zhao2024recommender,wu2024survey,geng2022recommendation,liu2025onerec,he2025plum} reformulates retrieval and ranking as sequence generation over semantic identifiers.
Most LLM-based work targets retrieval and early ranking; the re-ranking stage, and especially reasoning-based re-ranking, has received less attention.
GR2~\citep{liang2026generative} establishes design principles for reasoning re-rankers, i.e., semantic-ID grounding, high-quality reasoning traces via rejection sampling, and a re-ranking-specific RL reward, and is the autoregressive model we convert.
In the search domain, reasoning-augmented re-rankers such as ReaRank~\citep{zhang2025rearank} optimize relevance via RL. Our work is orthogonal to these: we keep the reasoning-re-ranker formulation fixed and change the \emph{decoder}, asking how to retain the AR re-ranker's accuracy under parallel diffusion decoding.

\subsection{Block-Diffusion Language Models and Parallel Decoding}
Masked- and block-diffusion language models~\citep{austin2021structured,shi2024simplified,lou2024discrete,sahoo2024simple,ye2025dream,zhu2026llada,arriola2025interpolating,wu2025fast,arriola2025block,labs2025mercury} generate text by iteratively denoising masked positions, committing the most confident positions per step. They offer parallel decoding and a tunable accuracy--latency trade-off, but their decoding is task-agnostic and treats every output token identically. Our approach departs from this: rather than a task-agnostic decoder, we recover answer validity through conversion fine-tuning from an AR initialization, so the converted re-ranker emits valid permutations of the candidate set on its own.

\subsection{AR-to-Diffusion Conversion and RL for Diffusion LMs}
A growing body of work initializes diffusion language models from autoregressive checkpoints and adapts them with reinforcement learning, including TraceRL~\citep{wang2025revolutionizing} and \mbox{d1}~\citep{zhao2026d1}. We build on this conversion-plus-RL paradigm but add two task-specific ingredients: conversion fine-tuning that transfers the AR model's permutation-validity competence into the diffusion decoder, and an on-policy distillation stage that recovers the AR model's behavior before RL.

\subsection{On-policy and Generalized Knowledge Distillation}
On-policy distillation~\citep{lu2025onpolicydistillation,su2026data}, an on-policy instance of generalized knowledge distillation~\citep{hinton2015distilling,agarwal2024policy}, supervises a student on its own sampled outputs rather than a fixed dataset, reducing the train/inference mismatch of off-policy distillation. Our OPD stage instantiates this idea for diffusion re-ranking: the student samples under its real block-diffusion decoding distribution and the AR model provides dense per-token targets, directly addressing the off-policy gap introduced by conversion fine-tuning.

\section{Next Steps}
\label{section:nextsteps}

\textbf{Speculative decoding with a diffusion draft.} The conversion of AR GR2 into Diffusion-GR2 opens a natural path to \emph{speculative decoding}~\citep{leviathan2023fast,chen2023accelerating}, in which the two decoders play complementary roles: a fast block-diffusion \emph{draft} and the autoregressive \emph{verifier}. Because the AR verify pass is the final quality gate, the draft is free to be recklessly fast, e.g., an aggressively pruned Diffusion-GR2 with a large block size and a low confidence threshold for maximal decoding parallelism, which cleanly decouples speed (owned by the draft) from quality (owned by the verifier) and promises further latency and FLOP reductions at no loss in ranking accuracy.

\textbf{Adaptive, pivot-aligned block sizing.} A second direction makes the block size \emph{adaptive} rather than a fixed compromise between acceptance rate and parallelism. The diffusion draft optimistically decodes the entire remaining span in one shot; a single AR verify pass then locates the \emph{pivot}, i.e., the position where the draft first diverges from the AR distribution, and the first block is resized to end at that pivot, with the procedure repeated over the remainder until the sequence is complete. Easy sequences are then decoded with near-full-sequence parallelism, while hard ones degrade gracefully to standard fixed-block cost. The same idea extends to a dynamic confidence threshold (larger blocks on high-agreement regions, smaller where the draft is uncertain) and to a coarse-to-fine verifier (a lightweight model that cheaply localizes the pivot region before the full AR verify). These ingredients compose: a pruned Diffusion-GR2 draft with progressively shrinking, pivot-aligned blocks stacks the speedup from pruning on top of the adaptive parallelism.

\section{Conclusion}
\label{section:conclusion}

We presented Diffusion-GR2, a recipe for converting an autoregressive reasoning re-ranker into a faster block-diffusion model while giving up as little accuracy as possible. Framing the conversion as a speed--accuracy trade-off, we identified its structural cause, i.e., invalid rankings under independent parallel decoding, and addressed it with conversion fine-tuning, which transfers the AR model's permutation-validity competence and recovers most of the conversion gap. We then minimized the residual off-policy gap with on-policy distillation, reaching near-parity with the AR re-ranker on Amazon Beauty while block-parallel decoding raises decode throughput by $2.4$--$3.5\times$ at the reasoning output length, with a reinforcement-learning stage on top of OPD to recover the remaining margin. We hope Diffusion-GR2 is useful as a template for deploying reasoning re-rankers at lower latency without sacrificing the reasoning that makes them accurate.

\clearpage
\newpage
\beginappendix

\section{On-Policy Distillation: Details}
\label{appendix:opd}

We draw $G' = 4$ on-policy trajectories per prompt from the converted model, detach the AR teacher, and minimize the forward KL of \cref{eq:opd} over committed positions, weighting reasoning and answer tokens with the same decoupled weights used in conversion fine-tuning. Teacher logits are computed once per trajectory and cached. We use a constant learning rate of $1\times10^{-6}$, $128$ prompts per optimizer step, and a single epoch over the training prompts; OPD converges within roughly $1.5$k steps on Beauty.

\section{Reinforcement Learning: Details}
\label{appendix:rl}

We optimize \cref{eq:grpo} with group size $G = 8$, clip range $\varepsilon = 0.2$, and the conditional format-reward weight $\alpha = 0.2$ from GR2. Because the converted model emits valid permutations, $R_{\mathrm{rank}}$ in \cref{eq:rankreward} is well-defined on essentially every rollout. We initialize from the OPD checkpoint for the OPD\,$\rightarrow$\,RL recipe. RL is run for $600$ steps at a learning rate of $5\times10^{-7}$.

\section{Hyperparameters}
\label{appendix:hparams}

\begin{table}[h]
    \centering
    \begin{NiceTabular}{lc}
    \toprule
    Stage & Beauty \\
    \midrule
    Conversion FT epochs            & $3$ \\
    Block size                      & $32$ \\
    Denoising steps (eval)          & $64$ \\
    OPD on-policy samples / prompt  & $4$ \\
    OPD learning rate               & $1\!\times\!10^{-6}$ \\
    RL group size $G$               & $8$ \\
    RL learning rate                & $5\!\times\!10^{-7}$ \\
    \bottomrule
    \end{NiceTabular}
    \caption{Key hyperparameters for the Diffusion-GR2 conversion pipeline.}
    \label{table:hparams}
\end{table}

\section{Chat-Formatted Prompt Template}
\label{appendix:template}

We reuse the GR2 chat template without modification: a system message defining the expert persona and the re-ranking task; a user message presenting the SID-grounded purchase history and the full candidate set with title and category metadata; and an assistant message containing a step-by-step, SID-grounded reasoning trace followed by a structured JSON object with the reasoning explanation and the ranked recommendation list. The answer span, i.e., the ranked list inside the JSON object, is the span the converted model learns to denoise into a valid permutation of the candidate set.

\clearpage
\newpage
\bibliographystyle{assets/plainnat}
\bibliography{paper}

@article{liang2026generative,
  title={Generative Reasoning Re-ranker},
  author={Liang, Mingfu and Li, Yufei and Xu, Jay and Asadi, Kavosh and Liu, Xi and Gu, Shuo and Rangadurai, Kaushik and Shyu, Frank and Wang, Shuaiwen and Yang, Song and others},
  journal={arXiv preprint arXiv:2602.07774},
  year={2026}
}

@article{wang2025revolutionizing,
  title={Revolutionizing reinforcement learning framework for diffusion large language models},
  author={Wang, Yinjie and Yang, Ling and Li, Bowen and Tian, Ye and Shen, Ke and Wang, Mengdi},
  journal={arXiv preprint arXiv:2509.06949},
  year={2025}
}

@inproceedings{zhang2025rearank,
  title={Rearank: Reasoning re-ranking agent via reinforcement learning},
  author={Zhang, Le and Wang, Bo and Qiu, Xipeng and Reddy, Siva and Agrawal, Aishwarya},
  booktitle={Proceedings of the 2025 Conference on Empirical Methods in Natural Language Processing},
  pages={2458--2471},
  year={2025}
}

@article{lu2025onpolicydistillation,
  author = {Kevin Lu and Thinking Machines Lab},
  title = {On-Policy Distillation},
  journal = {Thinking Machines Lab: Connectionism},
  year = {2025},
  note = {https://thinkingmachines.ai/blog/on-policy-distillation},
  doi = {10.64434/tml.20251026},
}

@inproceedings{arriola2025block,
  title={Block diffusion: Interpolating between autoregressive and diffusion language models},
  author={Arriola, Marianne and Gokaslan, Aaron and Chiu, Justin and Yang, Zhihan and Qi, Zhixuan and Han, Jiaqi and Sahoo, Subham and Kuleshov, Volodymyr},
  booktitle={International Conference on Learning Representations},
  volume={2025},
  pages={50726--50753},
  year={2025}
}

@article{labs2025mercury,
  title={Mercury: Ultra-Fast Language Models Based on Diffusion},
  author={Labs, Inception and Khanna, Samar and Kharbanda, Siddhant and Li, Shufan and Varma, Harshit and Wang, Eric and Birnbaum, Sawyer and Luo, Ziyang and Miraoui, Yanis and Palrecha, Akash and others},
  journal={arXiv preprint arXiv:2506.17298},
  year={2025},
  url={https://www.inceptionlabs.ai/blog/introducing-mercury}
}

@article{radford2018improving,
  title={Improving language understanding by generative pre-training},
  author={Radford, Alec and Narasimhan, Karthik},
  year={2018},
  url={https://api.semanticscholar.org/CorpusID:49313245}
}

@article{radford2019language,
  title={Language models are unsupervised multitask learners},
  author={Radford, Alec and Wu, Jeff and Child, Rewon and Luan, David and Amodei, Dario and Sutskever, Ilya},
  year={2019},
  url={https://api.semanticscholar.org/CorpusID:160025533}
}

@article{wu2025fast,
  title={Fast-dllm v2: Efficient block-diffusion llm},
  author={Wu, Chengyue and Zhang, Hao and Xue, Shuchen and Diao, Shizhe and Fu, Yonggan and Liu, Zhijian and Molchanov, Pavlo and Luo, Ping and Han, Song and Xie, Enze},
  journal={arXiv preprint arXiv:2509.26328},
  year={2025}
}

@inproceedings{zhu2026llada,
  title={Llada 1.5: Variance-reduced preference optimization for large language diffusion models},
  author={Zhu, Fengqi and Wang, Rongzhen and Nie, Shen and Zhang, Xiaolu and Wu, Chunwei and Zhou, Jun and Lin, Yankai and Wen, Ji-Rong and Li, Chongxuan},
  booktitle={Proceedings of the 64th Annual Meeting of the Association for Computational Linguistics (Volume 1: Long Papers)},
  pages={11425--11460},
  year={2026}
}

@article{ye2025dream,
  title={Dream 7B: Diffusion Large Language Models},
  author={Ye, Jiacheng and Xie, Zhihui and Zheng, Lin and Gao, Jiahui and Wu, Zirui and Jiang, Xin and Li, Zhenguo and Kong, Lingpeng},
  journal={arXiv preprint arXiv:2508.15487},
  year={2025}
}

@article{su2026data,
  title={Data-Efficient Autoregressive-to-Diffusion Language Models via On-Policy Distillation},
  author={Su, Xingyu and Helwig, Jacob and Parashar, Shubham and Chagi, Atharv and Jotsna, Lakshmi and Zhi, Degui and Caverlee, James and Kalathil, Dileep and Ji, Shuiwang},
  journal={arXiv preprint arXiv:2606.06712},
  year={2026}
}

@inproceedings{
  arriola2025interpolating,
  title={Interpolating Autoregressive and Discrete Denoising Diffusion Language Models},
  author={Marianne Arriola and Aaron Gokaslan and Justin T Chiu and Jiaqi Han and Zhihan Yang and Zhixuan Qi and Subham Sekhar Sahoo and Volodymyr Kuleshov},
  booktitle={The Thirteenth International Conference on Learning Representations},
  year={2025},
  url={https://openreview.net/forum?id=tyEyYT267x}
}

@article{zhao2026d1,
  title={d1: Scaling reasoning in diffusion large language models via reinforcement learning},
  author={Zhao, Siyan and Gupta, Devaansh and Zheng, Qinqing and Grover, Aditya},
  journal={Advances in Neural Information Processing Systems},
  volume={38},
  pages={56729--56762},
  year={2026}
}

@article{shi2024simplified,
  title={Simplified and generalized masked diffusion for discrete data},
  author={Shi, Jiaxin and Han, Kehang and Wang, Zhe and Doucet, Arnaud and Titsias, Michalis},
  journal={Advances in neural information processing systems},
  volume={37},
  pages={103131--103167},
  year={2024}
}

@article{liu2025onerec,
  author       = {Zhanyu Liu and
                  Shiyao Wang and
                  Xingmei Wang and
                  Rongzhou Zhang and
                  Jiaxin Deng and
                  Honghui Bao and
                  Jinghao Zhang and
                  Wuchao Li and
                  Pengfei Zheng and
                  Xiangyu Wu and
                  Yifei Hu and
                  Qigen Hu and
                  Xinchen Luo and
                  Lejian Ren and
                  Zixing Zhang and
                  Qianqian Wang and
                  Kuo Cai and
                  Yunfan Wu and
                  Hongtao Cheng and
                  Zexuan Cheng and
                  Lu Ren and
                  Huanjie Wang and
                  Yi Su and
                  Ruiming Tang and
                  Kun Gai and
                  Guorui Zhou},
  title        = {OneRec-Think: In-Text Reasoning for Generative Recommendation},
  journal      = {CoRR},
  volume       = {abs/2510.11639},
  year         = {2025},
  url          = {https://doi.org/10.48550/arXiv.2510.11639},
  doi          = {10.48550/ARXIV.2510.11639},
  eprinttype    = {arXiv},
  eprint       = {2510.11639},
  bibsource    = {dblp computer science bibliography, https://dblp.org}
}

@article{yu2025dapo,
  title={Dapo: An open-source llm reinforcement learning system at scale},
  author={Yu, Qiying and Zhang, Zheng and Zhu, Ruofei and Yuan, Yufeng and Zuo, Xiaochen and Yue, Yu and Dai, Weinan and Fan, Tiantian and Liu, Gaohong and Liu, Lingjun and others},
  journal={arXiv preprint arXiv:2503.14476},
  year={2025}
}

@article{shao2024deepseekmath,
  title={Deepseekmath: Pushing the limits of mathematical reasoning in open language models},
  author={Shao, Zhihong and Wang, Peiyi and Zhu, Qihao and Xu, Runxin and Song, Junxiao and Bi, Xiao and Zhang, Haowei and Zhang, Mingchuan and Li, YK and Wu, Yang and others},
  journal={arXiv preprint arXiv:2402.03300},
  year={2024}
}

@inproceedings{geng2022recommendation,
  title={Recommendation as language processing (rlp): A unified pretrain, personalized prompt \& predict paradigm (p5)},
  author={Geng, Shijie and Liu, Shuchang and Fu, Zuohui and Ge, Yingqiang and Zhang, Yongfeng},
  booktitle={Proceedings of the 16th ACM Conference on Recommender Systems},
  pages={299--315},
  year={2022}
}

@article{hinton2015distilling,
  title={Distilling the Knowledge in a Neural Network},
  author={Hinton, Geoffrey},
  journal={arXiv preprint arXiv:1503.02531},
  year={2015}
}

@article{rajput2023recommender,
  title={Recommender systems with generative retrieval},
  author={Rajput, Shashank and Mehta, Nikhil and Singh, Anima and Hulikal Keshavan, Raghunandan and Vu, Trung and Heldt, Lukasz and Hong, Lichan and Tay, Yi and Tran, Vinh and Samost, Jonah and others},
  journal={Advances in Neural Information Processing Systems},
  volume={36},
  pages={10299--10315},
  year={2023}
}

@article{yang2025qwen3,
  title={Qwen3 technical report},
  author={Yang, An and Li, Anfeng and Yang, Baosong and Zhang, Beichen and Hui, Binyuan and Zheng, Bo and Yu, Bowen and Gao, Chang and Huang, Chengen and Lv, Chenxu and others},
  journal={arXiv preprint arXiv:2505.09388},
  year={2025}
}

@article{he2025plum,
  title={Plum: Adapting pre-trained language models for industrial-scale generative recommendations},
  author={He, Ruining and Heldt, Lukasz and Hong, Lichan and Keshavan, Raghunandan and Mao, Shifan and Mehta, Nikhil and Su, Zhengyang and Tsai, Alicia and Wang, Yueqi and Wang, Shao-Chuan and others},
  journal={arXiv preprint arXiv:2510.07784},
  year={2025}
}

@inproceedings{zhou2025efficiency,
  title={The efficiency vs. accuracy trade-off: Optimizing rag-enhanced llm recommender systems using multi-head early exit},
  author={Zhou, Huixue and Gu, Hengrui and Zhan, Zaifu and Liu, Xi and Zhou, Kaixiong and Xiao, Yongkang and Liang, Mingfu and Govindan, Srinivas Prasad and Chawla, Piyush and Yang, Jiyan and others},
  booktitle={Proceedings of the 63rd Annual Meeting of the Association for Computational Linguistics (Volume 1: Long Papers)},
  pages={26443--26458},
  year={2025}
}

@inproceedings{lee2022autoregressive,
  title={Autoregressive image generation using residual quantization},
  author={Lee, Doyup and Kim, Chiheon and Kim, Saehoon and Cho, Minsu and Han, Wook-Shin},
  booktitle={Proceedings of the IEEE/CVF conference on computer vision and pattern recognition},
  pages={11523--11532},
  year={2022}
}

@article{ouyang2022training,
  title={Training language models to follow instructions with human feedback},
  author={Ouyang, Long and Wu, Jeffrey and Jiang, Xu and Almeida, Diogo and Wainwright, Carroll and Mishkin, Pamela and Zhang, Chong and Agarwal, Sandhini and Slama, Katarina and Ray, Alex and others},
  journal={Advances in neural information processing systems},
  volume={35},
  pages={27730--27744},
  year={2022}
}

@article{guo2025deepseek,
  title={Deepseek-r1: Incentivizing reasoning capability in llms via reinforcement learning},
  author={Guo, Daya and Yang, Dejian and Zhang, Haowei and Song, Junxiao and Zhang, Ruoyu and Xu, Runxin and Zhu, Qihao and Ma, Shirong and Wang, Peiyi and Bi, Xiao and others},
  journal={arXiv preprint arXiv:2501.12948},
  year={2025}
}

@inproceedings{zhang2025reasonrec,
  title={Reasonrec: A reasoning-augmented multimodal agent for unified recommendation},
  author={Zhang, Yihua and Liu, Xi and Zeng, Xihuan and Liang, Mingfu and Yang, Jiyan and Jin, Rong and Chen, Wen-Yen and Han, Yiping and Ma, Hao and Long, Bo and others},
  booktitle={ICML 2025 Workshop on Programmatic Representations for Agent Learning},
  year={2025}
}

@article{zhao2024recommender,
  title={Recommender systems in the era of large language models (llms)},
  author={Zhao, Zihuai and Fan, Wenqi and Li, Jiatong and Liu, Yunqing and Mei, Xiaowei and Wang, Yiqi and Wen, Zhen and Wang, Fei and Zhao, Xiangyu and Tang, Jiliang and others},
  journal={IEEE Transactions on Knowledge and Data Engineering},
  volume={36},
  number={11},
  pages={6889--6907},
  year={2024},
  publisher={IEEE}
}

@article{wu2024survey,
  title={A survey on large language models for recommendation},
  author={Wu, Likang and Zheng, Zhi and Qiu, Zhaopeng and Wang, Hao and Gu, Hongchao and Shen, Tingjia and Qin, Chuan and Zhu, Chen and Zhu, Hengshu and Liu, Qi and others},
  journal={World Wide Web},
  volume={27},
  number={5},
  pages={60},
  year={2024},
  publisher={Springer}
}

@article{schulman2017proximal,
  title={Proximal policy optimization algorithms},
  author={Schulman, John and Wolski, Filip and Dhariwal, Prafulla and Radford, Alec and Klimov, Oleg},
  journal={arXiv preprint arXiv:1707.06347},
  year={2017}
}

@inproceedings{agarwal2024policy,
  title={On-policy distillation of language models: Learning from self-generated mistakes},
  author={Agarwal, Rishabh and Vieillard, Nino and Zhou, Yongchao and Stanczyk, Piotr and Ramos Garea, Sabela and Geist, Matthieu and Bachem, Olivier},
  booktitle={International Conference on Learning Representations},
  volume={2024},
  pages={21246--21263},
  year={2024}
}

@inproceedings{zheng2023judging,
  title={Judging {LLM}-as-a-judge with {MT}-bench and chatbot arena},
  author={Zheng, Lianmin and Chiang, Wei-Lin and Sheng, Ying and Zhuang, Siyuan and Wu, Zhanghao and Zhuang, Yonghao and Lin, Zi and Li, Zhuohan and Li, Dacheng and Xing, Eric P. and Zhang, Hao and Gonzalez, Joseph E. and Stoica, Ion},
  booktitle={Advances in Neural Information Processing Systems (NeurIPS)},
  year={2023}
}

@inproceedings{mcauley2015image,
  title={Image-based recommendations on styles and substitutes},
  author={McAuley, Julian and Targett, Christopher and Shi, Qinfeng and van den Hengel, Anton},
  booktitle={Proceedings of the 38th International ACM SIGIR Conference on Research and Development in Information Retrieval},
  year={2015}
}

@inproceedings{leviathan2023fast,
  title={Fast inference from transformers via speculative decoding},
  author={Leviathan, Yaniv and Kalman, Matan and Matias, Yossi},
  booktitle={International Conference on Machine Learning (ICML)},
  year={2023}
}

@article{chen2023accelerating,
  title={Accelerating large language model decoding with speculative sampling},
  author={Chen, Charlie and Borgeaud, Sebastian and Irving, Geoffrey and Lespiau, Jean-Baptiste and Sifre, Laurent and Jumper, John},
  journal={arXiv preprint arXiv:2302.01318},
  year={2023}
}

@inproceedings{austin2021structured,
  title={Structured denoising diffusion models in discrete state-spaces},
  author={Austin, Jacob and Johnson, Daniel D. and Ho, Jonathan and Tarlow, Daniel and van den Berg, Rianne},
  booktitle={Advances in Neural Information Processing Systems (NeurIPS)},
  year={2021}
}

@inproceedings{lou2024discrete,
  title={Discrete diffusion modeling by estimating the ratios of the data distribution},
  author={Lou, Aaron and Meng, Chenlin and Ermon, Stefano},
  booktitle={International Conference on Machine Learning (ICML)},
  year={2024}
}

@inproceedings{sahoo2024simple,
  title={Simple and effective masked diffusion language models},
  author={Sahoo, Subham Sekhar and Arriola, Marianne and Schiff, Yair and Gokaslan, Aaron and Marroquin, Edgar and Chiu, Justin T. and Rush, Alexander M. and Kuleshov, Volodymyr},
  booktitle={Advances in Neural Information Processing Systems (NeurIPS)},
  year={2024}
}

@inproceedings{ross2011reduction,
  title={A reduction of imitation learning and structured prediction to no-regret online learning},
  author={Ross, St{\'e}phane and Gordon, Geoffrey and Bagnell, Drew},
  booktitle={Proceedings of the Fourteenth International Conference on Artificial Intelligence and Statistics (AISTATS)},
  year={2011}
}

@inproceedings{kim2016sequence,
  title={Sequence-level knowledge distillation},
  author={Kim, Yoon and Rush, Alexander M.},
  booktitle={Proceedings of the 2016 Conference on Empirical Methods in Natural Language Processing (EMNLP)},
  year={2016}
}

@inproceedings{ranzato2016sequence,
  title={Sequence level training with recurrent neural networks},
  author={Ranzato, Marc'Aurelio and Chopra, Sumit and Auli, Michael and Zaremba, Wojciech},
  booktitle={International Conference on Learning Representations (ICLR)},
  year={2016}
}
\end{document}